\documentclass[publicdomain]{eptcs}
\providecommand{\event}{ACT 2019} 

\usepackage[utf8]{inputenc}
\usepackage[english]{babel}
\usepackage[T1]{fontenc}
\usepackage{amsmath,amssymb,amsthm}
\usepackage{hyperref}
\usepackage[numbers,sort&compress]{natbib}
\usepackage{lipsum}
\usepackage[all,2cell,cmtip]{xy}
\UseTwocells
\usepackage[svgnames]{xcolor}
\usepackage{multirow}
\usepackage{comment}
\usepackage{array}
\newcolumntype{C}[1]{>{\centering\arraybackslash}p{#1}}

\usepackage{tikz}
\usetikzlibrary{decorations.markings}
\usetikzlibrary{shapes.geometric}

\tikzstyle{arrowhead}=[-latex,draw=black,line width=2.000]
\tikzstyle{wbox}=[rectangle,fill=white,draw=black]
\tikzstyle{species}=[circle,fill=white,draw=black,scale=.75]
\definecolor{lblue}{rgb}{0,250,255}
\tikzstyle{transition}=[rectangle,fill=lblue,draw=black,scale=.75]
\tikzstyle{arrow}=[-,postaction={decorate},decoration={markings,mark=at position .7 with {\arrow{>}}},line width=1.00]
\tikzstyle{simple}=[-,draw=black,line width=1.000]
\tikzstyle{ar}=[->,shorten >=1.2pt]

\tikzstyle{bus}=[circle,fill=black,draw=black,scale=0.4]
\tikzstyle{equation}=[circle,draw=black, fill=none,scale=0.4]
\tikzstyle{none}=[circle, inner sep=0]
\tikzstyle{junction}=[circle,fill=black,scale=0.18]

\pgfdeclarelayer{edgelayer}
\pgfdeclarelayer{nodelayer}
\pgfsetlayers{edgelayer,nodelayer,main}

\newlength{\localh}
\newlength{\locald}
\newbox\mybox
\def\mpp#1#2{\scalebox{1}{\setbox\mybox\hbox{#2}\localh\ht\mybox\locald\dp\mybox\addtolength{\localh}{-\locald}\raisebox{-#1\localh}{\box\mybox}}}

\DeclareRobustCommand{\coprod}{\mathop{\text{\fakecoprod}}}
\newcommand{\fakecoprod}{%
	\sbox0{$\prod$}%
	\smash{\raisebox{\dimexpr.9625\depth-\dp0}{\scalebox{1}[-1]{$\prod$}}}%
	\vphantom{$\prod$}%
}

\newcommand{\<}{\langle}
\renewcommand{\>}{\rangle}

\newcommand{\from}{\leftarrow}

\newcommand{\id}{\mathrm{id}}

\newcommand{\degC}{\ ^\circ\rm{C}}

\newcommand{\PP}{\mathbf{P}}
\newcommand{\MM}{\mathbf{M}}
\renewcommand{\SS}{\mathbf{S}}

\newcommand{\Interfer}{\mathtt{Intfr}}
\newcommand{\Chassis}{\mathtt{Chassis}}
\newcommand{\Optics}{\mathtt{Optics}}
\newcommand{\Bath}{\mathtt{Bath}}
\newcommand{\BoxTemp}{\mathtt{Box}}
\renewcommand{\Box}{\mathtt{Box}}

\newcommand{\Lab}{\mathtt{Lab}}
\newcommand{\Mixer}{\mathtt{Mixer}}
\newcommand{\Res}{\mathtt{Resevoir}}
\newcommand{\Heater}{\mathtt{Heater}}
\newcommand{\TempSys}{\mathtt{TempSys}}
\newcommand{\LengthSys}{\mathtt{LengthSys}}
\newcommand{\Sensors}{\mathtt{Sensors}}
\newcommand{\Actuators}{\mathtt{Actuators}}
\newcommand{\LSIob}{\mathtt{LSI}}

\newcommand{\heat}{\mathtt{heat}}
\newcommand{\laser}{\mathtt{laser}}

\newcommand{\water}{\mathtt{H_2O}}
\newcommand{\focus}{\mathtt{focus}}
\newcommand{\mix}{\mathtt{mix}}
\newcommand{\temp}{\mathtt{temp}}

\newcommand{\fringe}{\mathtt{fringe}}
\newcommand{\intens}{\mathtt{intensity}}
\newcommand{\drive}{\mathtt{drive}}
\newcommand{\setPt}{\mathtt{setPt}}

\newcommand{\Rel}{\textsf{Rel}}

\newcommand{\NN}{\mathbb{N}}

\newcommand{\PortGraph}{\mathsf{PortGraph}}
\newcommand{\LSI}{\mathsf{LSI}}
\newcommand{\Prob}{\mathsf{Prob}}
\newcommand{\Stoch}{\mathsf{Stoch}}
\newcommand{\MeanTime}{\mathsf{MeanTime}}
\newcommand{\Rate}{\mathsf{Rate}}
\newcommand{\IType}{\mathcal{I}}

\newcommand{\Err}{\mathsf{Err}}
\newcommand{\Pt}{\mathsf{Pt}}

\newcommand{\supp}{\mathsf{supp}}
\newcommand{\aggr}{\mathsf{aggr}}
\newcommand{\comp}{\mathsf{comp}}
\newcommand{\forget}{\mathsf{forget}}

\newcommand{\norm}{\mathsf{norm}}
\newcommand{\inv}{\mathsf{invert}}
\newcommand{\Hist}{\mathsf{Hist}}
\newcommand{\data}{\mathsf{data}}
\newcommand{\mean}{\mathsf{mean}}

\title{Modeling Hierarchical System with Operads}
\author{
Spencer Breiner
\institute{NIST\\Gaithersburg, MD, USA}
\email{spencer.breiner@nist.gov}
\and
Blake Pollard
\institute{NIST\\Gaithersburg, MD, USA}
\email{blake.pollard@nist.gov}
\and
Eswaran Subrahmanian
\institute{Carnegie Mellon University \\
Pittsburgh, PA,USA}
\email{sub@cmu.edu}
\and
Olivier Marie-Rose
\institute{Prometheus Computing\\
	New Market, MD, USA}
\email{o.marie-rose@prometheuscomputing.com}
}
\def\titlerunning{Modeling Hierarchical System with Operads}
\def\authorrunning{S. Breiner, B. Pollard, E. Subrahmanian \& O. Marie-Rose}

\begin{document}
\maketitle

\begin{abstract}
	This paper applies operads and functorial semantics to address the problem of failure diagnosis in complex systems. We start with a concrete example, developing a hierarchical interaction model for the Length Scale Interferometer, a high-precision measurement system operated by the US National Institute of Standards and Technology. The model is expressed in terms of combinatorial/diagrammatic structures called port-graphs, and we explain how to extract an operad $\LSI$ from a collection of these diagrams. Next we show how functors to the operad of probabilities organize and constrain the relative probabilities of component failure in the system. Finally, we show how to extend the analysis from general component failure to specific failure modes.
\end{abstract}

\section{Introduction}

Hierarchical systems are ubiquitous in nature, in engineering and in commerce. We draw trees to represent anatomic features, component breakdowns and command structures. Hierarchies use abstraction barriers and separation of concerns to analyze and simplify complex problems into manageable parts.

However, to represent a system as a tree involves abstracting away the interactions between its branches. An alternative viewpoint, exemplified by models called port-graphs, emphasizes the leaves of the tree (components) and the way that they are wired together to form a larger system. This additional information is critical for system analysis, but such diagrams quickly become unwieldy as the number of components grows.

Operads provide a nice mathematical formalism for interpolating between these two viewpoints. Our goal in this paper is to demonstrate, in concrete terms, the use of operads to organize both qualitative and quantitative descriptions of hierarchical systems. To that end, we begin by modeling a specific real-world system, the Length Scale Interferometer (LSI), a device for precise length calibration operated by the US National Institute for Standards and Technology.

We begin with a brief sketch of the LSI, its purpose and its design, followed by an informal review of operads. Next, we construct a specific operad $\LSI$ based on the architecture of the LSI system and expressed in terms of combinatorial/diagrammatic structures called port-graphs. This forms the basis for a functorial analysis of failure diagnosis. First, we can consider the relative probabilities of component and sub-component failure as a functor $\LSI\to\Prob$. Finally, we show how to integrate component level probabilities with more specific information about specific failure modes. 

\section{The Length Scale Interferometer}

The Length Scale Interferometer (LSI) \cite{LSI} is a precision length-measurement system operated by the US National Institute for Standards and Technology (NIST). Customers from around the world send length scales, marked plates or rods ranging in size from a millimeter to a meter, to be calibrated at NIST's Gaithersburg, MD campus. Using laser interferometry, the device accurately measures lengths to better than one part in one million (0.7 $\mu$m error across a one meter length), and can resolve markings on a scale down to 0.1 $\mu$m.

More formally, a \emph{nominal length scale} is defined by two distances: the total span $D$ and the spacing $d$, where $N=\frac{D}{d}\in\NN$; for example, a typical meter stick with millimeter hatch marks would have $D=1~\rm{m}$ and $d=1~\rm{mm}$. The scale has $N+1$ hatch marks located a $0,d,2d,\ldots Nd=D$. For a real scale we can set our frame of reference $Y$ by the first mark on the scale, but each of the others will exhibit an error $\varepsilon_{i}$, and the purpose of the LSI system is to identify these errors.

The basic idea is simple enough. The scale is installed on a movable chassis, which also contains one mirror of an interferometer. The length of the laser's path changes as the scale and chassis move, allowing us to infer the scale's position from the fringe count of the interferometer.

A calibration starts from the initial mark at $y_0=0$ and records the current fringe count $f_0$ on the interferometer's readout. The machine scans to the next hatch mark $y_1$, identified by an optical system, and locks the position of the chassis. The new fringe reading $f_1$, along with the laser's wavelength $\lambda_0$ and the index of refraction $n$, determines the associated error
\[y_1=d+\varepsilon_1=\lambda_0\cdot n\cdot(f_1-f_0).\]
We do the same for each of the rest of the marks, noting that all distances and errors are calculated relative to $y_0$ (rather than $y_{i-1}$).

In practice, there are a number of complications, of which the most significant involve unavoidable fluctuations in the environment. Here we are concerned with two main effects. First, the index of refraction, which we use to calculate lengths and errors, depends on temperature.\footnote{ The index of refraction also depends on pressure and humidity, but we ignore these in the interest of simplicity.} Second, thermal expansion means that the relative positioning of system elements changes over time.

For example, the relative separation between (the first hatch mark of) the scale and the chassis' mirror will vary due to expansion of the metal that connects them. Similarly, the scale itself expands, so that the positions of the hatch marks and their errors will vary. Consequently, we should reformulate the goal of the LSI as measuring length scales \emph{at a standard temperature of } $20\degC$.


This variation has two practical implications for the LSI system. First, we must apply temperature-dependent corrections to the gross measurements of the system (fringe counts). This is necessary both to calculate the true lengths that were measured and to convert from these to the desired (temperature-corrected) values. Second, in order to minimize correction error, the LSI must maintain environmental values as close to the target as possible; to obtain the 0.7 $\mu$m accuracy mentioned above, the system must operate within $.02\degC$ of the target value.

\section{Operads}

A (colored) operad is a mathematical structure for representing hierarchical (tree-structured) composition and decomposition. The fundamental element of an operad is an \emph{operation} $f$; every operation has an \emph{arity} $m$, a tuple of \emph{input objects} $Y_1,\ldots,Y_m$ and a single \emph{output object} $X$, which we usually indicate by writing $\alpha:Y_1,\ldots,Y_m\to X$ (or more compactly, $\<Y_{i}\>\to X$). When the input elements are not relevant, we may simply write $f:X$

Given additional operations $g_{i}:Y_{i}$, operadic composition substitutes these for the ``dummy variables'' $Y_{i}$ to obtain a new operation $f(g_1,\ldots,g_m)$ (or $f\<g_{i}\>$). If each $g_{i}$ has $m_{i}$ inputs, then the new composite has $\sum_{i} m_{i}$. Operadic associativity guarantees a well-defined composite for more deeply nested terms. When it is unambiguous we may also write $f(g)$ for composition along a single branch, rather than $f(\id,\ldots,g,\ldots,\id)$. See \cite{LeinsterOperad} for a careful exposition.

We will be interested in a typed variant of Spivak's operad of wiring diagrams \cite{operad}. We begin with a set of \emph{interfaces} $\IType$, which represent all the channels of interaction that occur within our system. These include both physical interactions ($\heat$ flow)) and digital signals ($\temp$ measurements). Formally, we specify $\IType$ at the outset, although in practice it can be inferred from usage in system diagrams.

Given $\IType$, a \emph{boundary} is a set of \emph{ports} $P$ together with a map (often left implicit) $P\to \IType$. We draw an interface as a box with $|P|$-many terminals, each labeled by a type (distinguished, if necessary, by subscripts). For example, the bath used in the LSI's temperature regulation system has a boundary

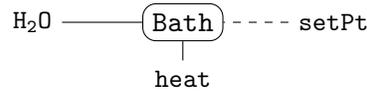
{\centering
	\begin{tikzpicture}
	\node[draw,rectangle,rounded corners=5pt] (tempsyst) at (0,0) {$\Bath$};
	\node (out2) at (0,-.75) {\small $\heat$};
	\node (out3) at (2,0) {\small $\setPt$};
	\node (out1) at (-2,0) {\small $\water$};
	\draw (out1) to (tempsyst);
	\draw (out2) to (tempsyst);
	\draw[dashed] (out3) to (tempsyst);
	\end{tikzpicture}.\\
}

This indicates that the bath has three main interactions: heat flow to the length measurement enclosure, chilled water provided from an outside system and a data stream that modifies the set point of an internal heating controller. For now our labels are just place-holders, but they will be refined as we elaborate the model.

In the $\PortGraph$ operad, an operation $\alpha:\<Q_{i}\>\to P$ is a system architecture, modeled as a port-graph, in which $Q_{i}$ are sub-component boundaries and $P$ is an external system boundary. The ports in $P$ and $Q_{i}$ can be connected via wires, which are also typed, and more properly described as ``hyper-wires'' given that a single wire can connect more than two ports.

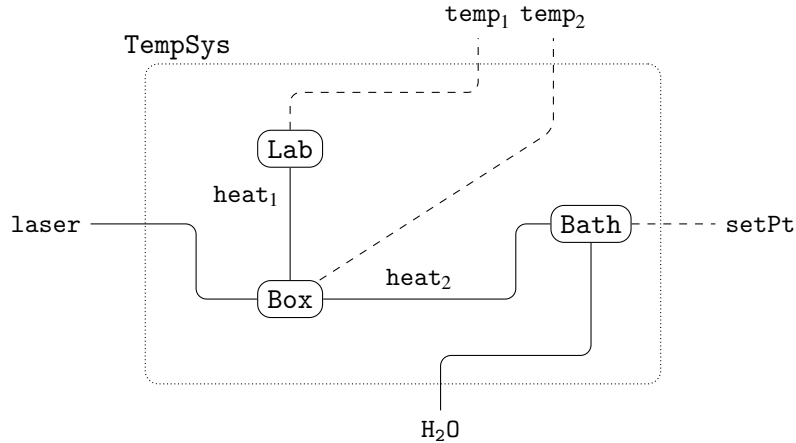
\begin{figure}
	\begin{center}
		\begin{tikzpicture}
		\node[draw,rectangle,rounded corners=5pt] (room) at (-2,1) {$\Lab$};
		\node[draw,rectangle,rounded corners=5pt] (box) at (-2,-1) {$\BoxTemp$};
		\node[draw,rectangle,rounded corners=5pt] (bath) at (2,0) {$\Bath$};
		\node[draw,rectangle,rounded corners=5pt,densely dotted] (target) at (-.5,0) {\rule{0cm}{4cm} \rule{6.5cm}{0cm}};
		\node (targetlabel) at (-3.5,2.35) {$\TempSys$}; 
		\node (data1) at (.5,2.75) {\small $\temp_1$};
		\node (data2) at (1.5,2.75) {\small $\temp_2$};
		\node (set) at (4.25,0) {\small$\setPt$};
		\node (ref) at (-5.25,0) {\small$\laser$};
		\node (water) at (0,-2.75) {\small$\water$};
		\draw[rounded corners,dashed] (room.north) to (-2,1.75) to (.5,1.75) to (data1);
		\draw[rounded corners,dashed] (box) to (1.5,1.25) to (data2);
		\draw[rounded corners] (box) to (-3.25,-1) to (-3.25,0) to (ref);
		\draw (room) to node[left, near start] {\small$\heat_1$} (box);
		\draw[rounded corners] (box) to node[above] {\small$\heat_2$} (1,-1) to (1,0) to (bath);
		\draw[rounded corners,dashed] (bath) to (set);
		\draw[rounded corners] (bath) to (2,-1.75) to (0,-1.75) to (water);
		\end{tikzpicture}
	\end{center}
	\caption{A schematic decomposition of the LSI's temperature regulation system.}
	\label{fig:TempSys}
\end{figure}

For example the LSI's temperature control system, shown in Figure \ref{fig:TempSys}, has three subsystems: one controls the ambient lab temperature ($\pm 0.2\degC$), another measures the internal temperature of the measurement enclosure ($\pm .005\degC$) and a third manages the chilled water bath used to control the system's temperature.

Physically, the bath and the lab environment both impact the enclosure through heat flow and this, in turn, affects the length measurement subsystem via refraction of the laser. Another physical input to the system, a chilled water source, is a shared resource controlled outside of the lab.

There are also digital flows involved in the system. Two streams of temperature data are produced by the lab and the enclosure. There is also a control interaction, with a variable set point for a heating element in the bath. Diagrammatically, it can be useful to distinguish between physical and informational flows; for example, the latter have only indirect effects on the temperature of the system, in contrast to the physical inputs. Formally, the use of different notations for different interaction types can be justified by ``typing the types'' via a function $\IType \to \{\mathrm{physical},\mathrm{digital}\}$.

We can succinctly represent a port-graph architecture $\alpha$ as a zig-zag of functions, two of which commute over types:
\begin{equation}
\vcenter{\xymatrix{
		C & Q \ar[l] \ar[r] \ar[rd] & W \ar[d] & P \ar[l] \ar[dl]\\
		& & \IType & \\
}}
\label{eq:arch}
\end{equation}
Here $C=\comp(\alpha)\cong\{1,\ldots,n\}$ is the set of (black-box) components in the architecture, $Q=\coprod_{i} Q_{i}$ is the set of internal ports, $W$ is the set of wires and $P$ is the set of external ports. Both ports and wires are typed, and the resulting triangles should commute.

Operadic composition acts by substituting one architecture into another. For example, Figure \ref{fig:subst} shows the result of substituting a lower-level architecture composed of a mixer, a resevoir and a heater for the $\Bath$ component from Figure \ref{fig:TempSys}.

In the general case we may substitute for all $Q_{i}$ simultaneously, so suppose that we have a decomposition (suppressing types) ${D_{i} \from R_{i} \to V_{i} \from Q_{i}}$ for each top-level component $i\in C$. Clearly the external ports $P$ are unchanged by substitution inside $Q_{i}$. The set of internal boundaries is given by $D=\coprod_{i} D_{i}$; similarly the internal ports of the architecture will be $R=\coprod_{i} R_{i}$.

Finally, we can compute the wire set for the aggregate by identifying an outer wire $w$ with an inner wire $v_{i}$ whenever they share an intermediate port $Q_{i}$. Formally, this corresponds to a pushout, and typing on the new wires exists by virtue of the associated mapping property:
\begin{equation}
\vcenter{\xymatrix@=2ex{
		&&& V\underset{Q}{+}W &\\
		\coprod_{i} D_{i} & \coprod_{i} R_{i} \ar[l] \ar[r] & \coprod_{i} V_{i} \ar[ur] && W \ar[ul] & P \ar[l]\\
		&&& \coprod_{i} Q_{i} \ar[ur] \ar[ul] & \\
}}
\label{eq:arch_comp}
\end{equation}
Equations \ref{eq:arch} and \ref{eq:arch_comp} (along with the set $\IType$) define an operad $\PortGraph$.

\begin{figure}
	\begin{center}
		\scalebox{.9}{
			\begin{tikzpicture}
			\node[draw,rectangle,rounded corners=5pt] (room) at (-3,1) {$\Lab$};
			\node[draw,rectangle,rounded corners=5pt] (box) at (-3,-1) {$\BoxTemp$};
			\node[draw,rectangle,rounded corners=5pt] (mix) at (.75,.5) {$\Mixer$};
			\node[draw,rectangle,rounded corners=5pt] (res) at (.75,-1) {$\Res$};
			\node[draw,rectangle,rounded corners=5pt] (omega) at (4,-.25) {$\Heater$};
			\node[draw,rectangle,rounded corners=5pt,densely dotted] (target) at (.25,0) {\rule{0cm}{4cm} \rule{10.5cm}{0cm}};
			\node (targetlabel) at (-4.5,2.4) {$\TempSys$}; 
			\node[draw,rectangle,rounded corners=5pt,densely dotted] (target2) at (2.5,-.25) {\rule{0cm}{2.5cm} \rule{5.5cm}{0cm}};
			\node (target2label) at (5,1.4) {$\Bath$};
			\node (data1) at (-1.5,2.75) {\small $\temp_1$};
			\node (data2) at (-.5,2.75) {\small $\temp_2$};
			\node (set) at (7,-.25) {\small $\setPt$};
			\node (ref) at (-6.5,-.25) {\small $\laser$};
			\node (water) at (.75,-2.75) {\small $\water$};
			\node[draw,circle] (mixdot) at (mix|-omega) {};
			\draw[rounded corners,dashed] (room.north) to (-3,1.75) to (-1.5,1.75) to (data1);
			\draw[rounded corners,dashed] (box) to (-.5,1.5) to (data2);
			\draw[rounded corners] (box) to (-4.25,-1) to (-4.25,-.25) to (ref);
			\draw (room) to node[left] {\small $\heat_1$} (box);
			\draw[rounded corners] (mix) to (mixdot);
			\draw[rounded corners] (res) to (mixdot);
			\draw[rounded corners] (mixdot) to node[above,near start] {$\heat_3$} (omega);
			\draw[rounded corners] (box) to node[below] {\small $\heat_2$} (res);
			\draw[rounded corners,dashed] (omega) to (set);
			\draw[rounded corners] (res) to (water);
			\draw[dashed,rounded corners] (res) to node[below] {\small$\temp$} (res-|omega) to (omega);
			\end{tikzpicture}
		}
	\end{center}
	\caption{
		An operadic composition $\tau(\beta)$ for two architectures $\tau:\Bath,\Lab,\BoxTemp\to\TempSys$ and $\beta:\Mixer,\Res,\Heater\to\Bath$.
	}
	\label{fig:subst}
\end{figure}

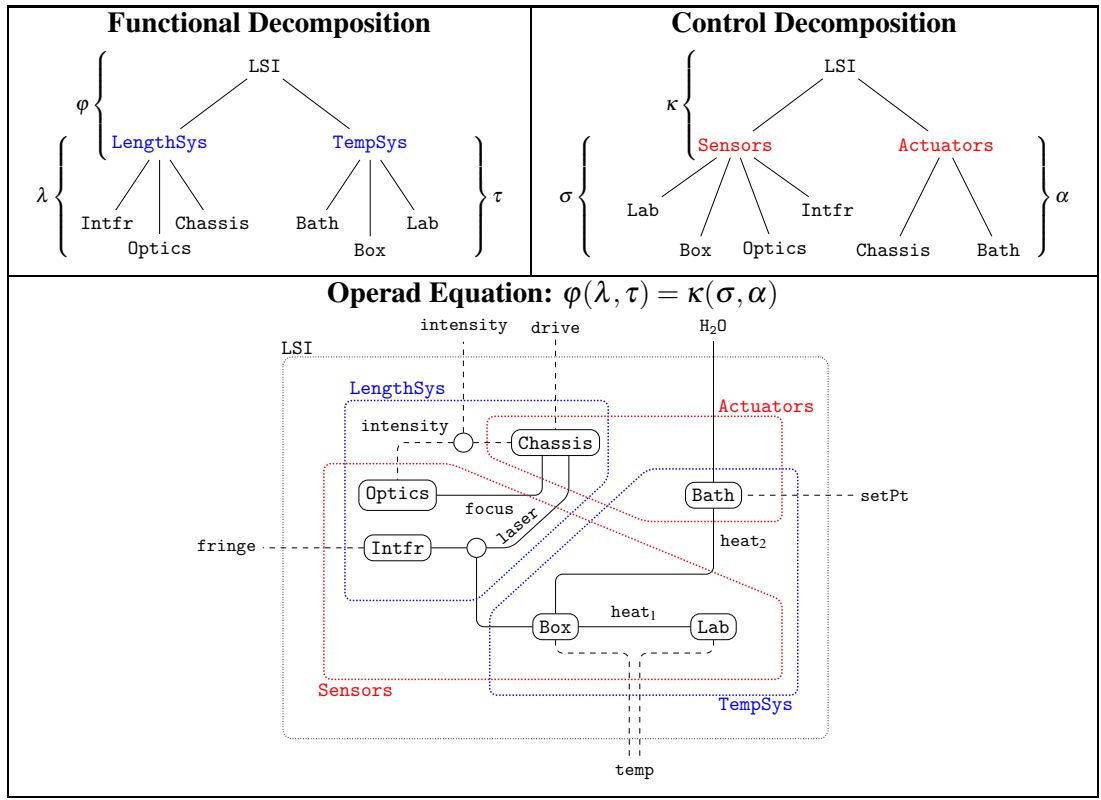
\begin{figure}[t]
	\begin{center}
		\begin{tabular}{|c|c|}
			\hline
			\textbf{Functional Decomposition} & \textbf{Control Decomposition}\\
			\scalebox{.7}{
				\begin{tikzpicture}
				\node (1) at (3,0) {$\Lab$};
				\node (2) at (2,-.5) {$\BoxTemp$};
				\node (3) at (1,0) {$\Bath$};
				\node (4) at (-1,0) {$\Chassis$};
				\node (5) at (-2,-.5) {$\Optics$};
				\node (6) at (-3,0) {$\Interfer$};
				\node[blue] (11) at (2,1.5) {$\TempSys$};
				\node[blue] (12) at (-2,1.5) {$\LengthSys$};
				\node (0) at (0,3) {$\LSIob$};
				\draw (1) to (11);
				\draw (2) to (11);
				\draw (3) to (11);
				\draw (4) to (12);
				\draw (5) to (12);
				\draw (6) to (12);
				\draw (11) to (0);
				\draw (12) to (0);
				\node (f) at (-3.25,2.25) {$\varphi \left\{\rule{0cm}{1.155cm}\right.$};
				\node (ell) at (-4,.5) {$\lambda \left\{\rule{0cm}{1.25cm}\right.$};
				\node (ell) at (4.2,.5) {$\left.\rule{0cm}{1.25cm}\right\}\tau$};
				\end{tikzpicture}
			}
			&
			\scalebox{.7}{
				\begin{tikzpicture}
				\node (1) at (-3.75,0.25) {$\Lab$};
				\node (2) at (-2.75,-.5) {$\BoxTemp$};
				\node (3) at (-1.25,-.5) {$\Optics$};
				\node (4) at (-.25,0.25) {$\Interfer$};
				\node (5) at (1,-.5) {$\Chassis$};
				\node (6) at (3,-.5) {$\Bath$};
				\node[red] (11) at (-2,1.5) {$\Sensors$};
				\node[red] (12) at (2,1.5) {$\Actuators$};
				\node (0) at (0,3) {$\LSIob$};
				\draw (1) to (11);
				\draw (2) to (11);
				\draw (3) to (11);
				\draw (4) to (11);
				\draw (5) to (12);
				\draw (6) to (12);
				\draw (11) to (0);
				\draw (12) to (0);
				\node (c) at (-3,2.25) {$\kappa \left\{\rule{0cm}{1.15cm}\right.$};
				\node (s) at (-5,.5) {$\sigma \left\{\rule{0cm}{1.25cm}\right.$};
				\node (a) at (4,.5) {$\left.\rule{0cm}{1.25cm}\right\}\alpha$};
				\end{tikzpicture}
			}\\\hline
			\multicolumn{2}{|c|}{\textbf{Operad Equation:} $\varphi(\lambda,\tau)=\kappa(\sigma,\alpha)$}\\
			\multicolumn{2}{|c|}{
				\scalebox{.7}{
					\begin{tikzpicture}
					\node[draw,rectangle,rounded corners=5pt] (optics) at (-3,1) {$\Optics$};
					\node[draw,rectangle,rounded corners=5pt] (chassis) at (0,2) {$\Chassis$};
					\node[draw,rectangle,rounded corners=5pt] (laser) at (-3,0) {$\Interfer$};
					\node[draw,rectangle,rounded corners=5pt] (roomtemp) at (3,-1.5) {$\Lab$};
					\node[draw,rectangle,rounded corners=5pt] (boxtemp) at (0,-1.5) {$\BoxTemp$};
					\node[draw,rectangle,rounded corners=5pt] (bath) at (3,1) {$\Bath$};
					\node[draw,rectangle,rounded corners=5pt,densely dotted] (target0) at (0,0) {\rule{0cm}{7cm} \rule{10cm}{0cm}};
					\node (target0label) at (-4.9,3.8) {$\LSIob$}; 
					\node[draw,circle] (laserdot) at (-1.5,0) {};
					\draw (laser) to (laserdot);
					\draw[rounded corners] (chassis.-45) to (chassis.-45|-optics) to node[above, near end,rotate=40] {\small $\laser$} (chassis.west|-laserdot) to (laserdot);
					\draw[rounded corners] (boxtemp.180) to (boxtemp-|laserdot) to (laserdot);
					\node[draw,circle] (intensdot) at (-1.75,2) {};
					\node (intensity) at (-1.75,4.2) {\small $\intens$};
					\draw[dashed,rounded corners] (intensdot) to node[above,very near end] {\small $\intens$} (chassis-|optics) to (optics);
					\draw[dashed,rounded corners] (intensdot) to (intensity);
					\draw[dashed,rounded corners] (intensdot) to (chassis);
					\draw[rounded corners] (chassis.-135) to (chassis.-135|-optics) to node[below] {\small $\focus$} (optics);
					\node (position) at (0,4.2) {\small $\drive$};
					\draw[dashed] (position) to (chassis);
					\node (fringes) at (-6.25,0) {\small $\fringe$};
					\draw[dashed] (laser) to (fringes);
					\draw[rounded corners] (boxtemp) to node[above] {\small $\heat_1$} (roomtemp);
					\draw[rounded corners] (boxtemp) to (0,-.5) to (3,-.5) to node[right] {\small$\heat_2$} (bath);
					\node (set) at (6.25,1) {\small $\setPt$};
					\draw[dashed] (set) to (bath);
					\node (h20) at (3,4.2) {\small $\water$};
					\draw (h20) to (bath);
					\node (tempSignal) at (1.5,-4.25) {\small $\temp$};
					\draw[dashed, rounded corners] (roomtemp) to (3,-2) to (1.6,-2) to (1.6,-4);
					\draw[dashed, rounded corners] (boxtemp) to (0,-2) to (1.4,-2) to (1.4,-4);
					\node[blue] (target1label) at (3.8,-3) {$\TempSys$}; 
					\draw[densely dotted,thick,rounded corners,blue] (4.6,0) to (4.6,1.5) to (1.75,1.5) to (-1.25,-1.25) to (-1.25,-2.8) to (4.6,-2.8) to (4.6,0);
					\node[blue] (target2label) at (-3,3) {$\LengthSys$}; 
					\draw[densely dotted,thick,rounded corners,blue] (-4,0) to (-4,2.8) to (1,2.8) to (1,1.3) to (-1.5,-1) to (-4,-1) to (-4,0);
					\draw[densely dotted,thick,rounded corners,red] (-4.4,0) to (-4.4,1.6) to (-2,1.6) to (4.3,-1) to (4.3,-2.5) to (-4.4,-2.5) to (-4.4,0);
					\node[red] (target1label) at (-3.8,-2.7) {$\Sensors$}; 
					\draw[thick,densely dotted,rounded corners,red] (2,2.5) to (4.3,2.5) to (4.3,.5) to (1.7,.5) to (-1.3,1.75) to (-1.3,2.5) to (2,2.5);
					\node[red] (target2label) at (4,2.7) {$\Actuators$}; 
					\end{tikzpicture}
				}
			}\\\hline
		\end{tabular}
	\end{center}
	\caption{Operad equations represent common refinements between distinct hierarchies.}
	\label{fig:decomps}
\end{figure}

One nice feature of operads is that, although they allow us to represent hierarchical information, they do not constrain us to work within a single hierarchy. In particular, different perspectives often suggest different decompositions for a system, and we can make sense of these distinctions using equations in the operad.

An example is shown in Figure \ref{fig:decomps}. On one hand we can decompose the LSI system functionally (blue boundaries), separating the elements involved in length measurement from those used for temperature control. The second shows a more generic, control-theoretic perspective (red boundaries), separating out components that produce observations (sensors) from those which modify the state of the system (actuators). We could also consider a ``geographic'' decomposition based on the physical access to the components (e.g., inside/outside the system enclosure). An equation in the $\PortGraph$ operad indicates that two incompatible hierarchies (e.g., function vs. control) have a common refinement at a lower level of abstraction.

On first encounter, it can be easy to mix up port graph diagrams, which represent specific architectures, with the $\PortGraph$ operad, which represents \emph{all possible} architectures. A specific system, as represented by a collection of diagrams, can be compiled into a ``sub-operad'' (faithful functor) $\LSI\subseteq\PortGraph$ involving only the boundaries and architectures that appear in our diagrams.\footnote{However, note that diagrams with nested interfaces encode multiple architectures; six operations ($\varphi$, $\lambda$, $\tau$, $\kappa$, $\sigma$ and $\alpha$) and an equation can be extracted from the single diagram in Figure \ref{fig:decomps}.} A complete description of the $\LSI$ operad, compiled from Figures \ref{fig:TempSys}, \ref{fig:subst} and \ref{fig:decomps}, is given in the Appendix, Tables \ref{tab:LSI1} \& \ref{tab:LSI2}.

\section{Component failure}
\label{sec:fail}

In the last section we constructed an operad $\LSI$ whose operations are hierarchically organized architectures of the Length Scale Interferometer. Once we have laid out the architecture of the system, we can use it to structure system analyses.

\begin{table}[b!]
	\centering
	\begin{tabular}{|c|ccc||c|ccc|}
		\multicolumn{8}{c}{\textbf{LSI Component Failure Model}}\\\hline
		\multirow{2}{*}{$\varphi(ls,ts)$} & $ls$&$\mapsto$& 40\%
		& \multirow{2}{*}{$\kappa(sn,ac)$} & $sn$ & $\mapsto$ & 28\%\\
		& $ts$&$\mapsto$& 60\%
		&& $ac$ & $\mapsto$ & 72\%
		\\\hline
		\multirow{3}{*}{$\lambda(in,op,ch)$} & $in$ & $\mapsto$ & 10\%
		& \multirow{4}{*}{$\sigma(lb,bt,op,in)$} & $lb$ & $\mapsto$ & 21.4\%\\
		& $op$ & $\mapsto$ & 30\%
		& & $bt$ & $\mapsto$ & 21.4\%\\
		& $ch$ & $\mapsto$ & 60\%
		& & $op$ & $\mapsto$ & 42.9\%\\\cline{1-4}
		\multirow{3}{*}{$\tau(ba,bx,lb)$} & $ba$ & $\mapsto$ & 80\%
		& & $in$ & $\mapsto$ & 14.3\%\\\cline{5-8}
		& $bx$ & $\mapsto$ & 10\%
		& \multirow{2}{*}{$\alpha(ch,ba)$} & $ch$ & $\mapsto$ & 33.3\%\\
		& $lb$ & $\mapsto$ & 10\%
		& & $ba$ & $\mapsto$ & 66.7\%\\\cline{1-8}
		\multirow{3}{*}{$\beta(ht,mx,rs)$}
		& $ht$ & $\mapsto$ & 50\% \\
		& $mx$ & $\mapsto$ & 30\% \\
		& $rs$ & $\mapsto$ & 20\% \\
		\cline{1-4}
	\end{tabular}
	
	\caption{A probabilistic model of component failure presented as an operad functor $\LSI\to\Prob$.}
	\label{tab:fail}
\end{table}

The general principle of functorial semantics is that a model can be regarded as a mapping (called a functor) from architectural (syntactic) elements to computational (semantic) ones, but that this mapping must be assigned in a way that respects the compositional structure of the syntax. Probability provides a good example.

A very simple model of component failure might conclude, based on historical data, that a failure in the LSI will be located in the temperature regulation subsystem $60\%$ of the time, and in the length measurement subsystem the remaining $40\%$ of the time. The probability distribution $p=(ls\mapsto 40\%,ts\mapsto 60\%)$ is itself an operation in an operad, and we can think of the failure model as a mapping from the functional architecture $\varphi$ to the distribution $p$.

Suppose, moreover, that we also know the probabilities of failures within the temperature regulation system, which can involve the bath ($ba\mapsto 80\%$), the lab temperature system ($lb\mapsto 10\%$) or enclosure (box) temperature measurement system ($bx\mapsto 10\%$). Relative probabilities compose by multiplication (conditioning), so that given a fault somewhere in the LSI, there is a $80\%\times 60\%=48\%$ chance that it involves the bath. This is an instance of functoriality: composite architectures map to composite distributions.

More precisely, there is an operad of probabilities called $\Prob$. $\Prob$ is a ``plain'' operad, so that each operation has a fixed number of inputs (arity) $m$ (in contrast to $\PortGraph$, where each operation $\pi$ has a ``shape'' $\<Q_{i}\>\to P$). An operation $\Prob$ is just a finite probability distribution $p=\<i\mapsto p_{i}\>_{i<m}$, and its arity is the size of the index set $m$. The operadic (tree-structured) composition of $p$ with $m$-many additional distributions $q_{i}=\<j\mapsto q_{ij}\>_{j<m_{i}}$ is defined by 
\begin{equation}
p(q_1,\ldots,q_n) = \Big\<(i,j)\mapsto p_{i}\cdot q_{ij}\Big\>, 
\end{equation}
where the index set is the collection of pairs $(i<m,j<m_{i})$.

\begin{table}[b]
	\centering\framebox{
		$\begin{array}{ccccccccc}
		in:\Interfer & & \mapsto & & \overbrace{40\%}^{\varphi}\times \overbrace{10\%}^{\lambda} & = & 4\% & = & \overbrace{28\%}^{\kappa} \times \overbrace{14.3\%}^{\sigma}\\
		op:\Optics & & \mapsto & & 40\%\times 30\% & = & 12\% & = & 28\%\times42.9\%\\
		ch:\Chassis & & \mapsto & & 40\%\times 60\% & = & 24\% & = & 72\%\times \overbrace{33.3\%}^{\alpha}\\
		ba:\Bath & & \mapsto & & 60\%\times \overbrace{80\%}^{\tau} & = & 48\%  & = & 72\%\times 66.7\%\\
		bt:\BoxTemp & & \mapsto & & 60\%\times 10\% & = & 6\% & = & 28\%\times \overbrace{21.4\%}^{\sigma}\\
		rt:\Lab & & \mapsto & & 60\%\times 10\% & = & 6\% & = & 28\%\times 21.4\%\\
		\end{array}$}
	\caption{Functorial coherence equations induced by the equation $\varphi(\lambda,\tau)=\kappa(\alpha,\sigma)$}
	\label{tab:probfunctor}
\end{table}

This allows us to think of the simple failure model described above as a functor $\PP:\LSI\to\Prob$. The data associated with a specific instance is shown in Table \ref{tab:fail}. It sends each architecture $Q_1,\ldots,Q_m\to P$ to a probability distribution on $m$ elements, which should be interpreted as the relative probability of a failure in $Q_{i}$, \emph{given a failure in} $P$.

In order to define a functor, these probability assignments should satisfy the $\LSI$'s composition equation (Figure \ref{fig:decomps}). This defines some global constraints on how probability can be apportioned between different systems. For example, from above we know that there $80\%\times 60\%=48\%$ chance that a given failure occurs in the $\Bath$. If we also know that problems with actuators are responsible for 72\% of all faults in the LSI ($\kappa:ac\mapsto72\%$), then we can infer that the bath is twice as likely to malfunction as the chassis (third and fourth equations of Table \ref{tab:probfunctor}). All in all, the $\LSI$'s coherence equation $\varphi(\lambda,\tau)=\kappa(\sigma,\alpha)$ generates six probability equations, corresponding to the six components involved in the composed architecture.

Even in the absence of data we can represent these equations symbolically--the top line corresponds to $\varphi(ls)\cdot\lambda(in)=\kappa(sn)\cdot\sigma(in)$--and in this form we can view operadic coherence equations as data integrity constraints or rules of inference.

\section{Requirements and behavior}
\label{sec:failmode}

Though the operadic structure of the LSI model is useful for organizing failure probabilities, our analysis so far is deficient in that it considers only the components that fail, and not what goes wrong with them. In this section we consider more specific failure modes for the LSI, and their analysis using the classical method of fault trees.

Failure modes are best understood in the context of requirements. For example, the key functional requirement of the $\TempSys$ subsystem is that the laser's temperature should stay within $0.02$ degrees of $20\degC$. Correspondingly, we have two failure modes
\begin{equation}\label{eq:tempreq}\begin{array}{C{5cm}C{5cm}}
$T_\laser<19.98\degC,$ & $20.02\degC<T_\laser$.\\
\end{array}\end{equation}
In a similar fashion, we can associate a set of failure modes $\Err(P)$ with each boundary $P\in\LSI$. Typically these should reference features of the associated boundary (i.e., $\laser$ is an element of the boundary $\TempSys$). 

If $T_\laser$ is low, this may happen for a number of reasons. Either the $\Bath$ or the ambient $\Lab$ might be too cold, or the insulation of the $\Box$ might be leaking. However, we can say with confidence that a \emph{high} $\Bath$ temperature is not the cause. For each failure mode at the lower level we ask whether or not it can lead to a high temp on $\laser$. Repeating this procedure for each high-level failure, the ``can cause'' relation defines a functor $\MM:\LSI\to\Rel^+$.

Here $\Rel^+$ is an operad of relations, in which the objects are sets, and an arrow $\<Y_{i}\>\to X$ is a relation
\[R\subseteq \left(\coprod_{i} X_{i}\right)\times Y \cong \prod_{i}\left(X_{i}\times Y\right),\]
or, equivalently, a family of relations $R_{i}\subseteq X_{i}\times Y$. With the usual composition of relations, functoriality asserts the transitivity of causation: if a malfunctioning heater can cause a cold bath, and a cold bath can cause a low laser temp, then a malfunctioning heater can cause a low laser temp.

\section{Synthesizing semantics}
\label{sec:synth}

Intuitively, there should be some relationship between the failure probabilities from section \ref{sec:fail} and the failure modes of the last section. At a minimum, if some component $Y_{i}$ does not cause \emph{any} errors in $X$, then the associated probability should be zero. In this section we formalize this relationship functorially, as a joint lifting of $\PP$ and $\MM$:
\begin{equation}\label{eq:lifting}\vcenter{\xymatrix{
		& \LSI \ar[dl]_{\PP} \ar[rd]^{\MM} \ar@{-->}[d] & \\
		\Prob & \bullet \ar@{-->}[l] \ar@{-->}[r] & \Rel^+ \\
}}\end{equation}

Given sets $X$ and $Y$, a (stochastic) kernel $p:X\leadsto Y$ is an $X$-indexed set of probability distributions $p_x:Y\to[0,1]$. We think of $p$ as a stochastic mapping, which we emphasize by writing $p_x(y)=p(x\mapsto y)$. There is a category $\Stoch$ in which the objects are sets and two kernels $p:X\leadsto Y$ and $q:Y\leadsto Z$ compose by marginalization: 
\[p.q:\ (x\mapsto z)\longmapsto \sum_{y\in Y} p(x\mapsto y)\cdot q(y\mapsto z).\]
Using essentially the same technique we can construct an operad $\Stoch^+$ in which an arrow $p:\<Y_{i}\>\to X$ is a kernel $X\leadsto \coprod_{i} Y_{i}$.

Every kernel defines a relation---its support $\{(x,y)\ |\ p(x\mapsto y)>0\}$---providing a functor $\supp:\Stoch^+\to\Rel^+$. However, $\Stoch^+$ is not good enough to provide a joint lifting, as it does not provide a functor to $\Prob$. In Bayesian terms, a kernel represents a conditional probability, and to generate an absolute probability we need to combine this with a prior.

A single distribution over $X$ can be represented as a kernel ${r:1\leadsto X}$, also called a \emph{point} in $\Stoch$. There is a category $\Pt(\Stoch)$ in which the objects are pairs $\<X,r\>$ and an arrow $\<Y,s\>\to\<X,r\>$ is a kernel $p:Y\leadsto X$ forming a commutative triangle
\begin{equation}\label{eq:ptmap}\vcenter{\xymatrix{
		& 1 \ar@{~>}[ld]_{s} \ar@{~>}[rd]^{r} &\\
		Y \ar@{~>}[rr]_{p} && X \\
}}\end{equation}

Again, we can use coproducts to define an associated operad $\Pt(\Stoch)^+$. As above, an arrow $\<Y_{i},s_{i}\>\to\<X,r\>$ is a kernel $p:X\leadsto\coprod_{i} Y_{i}$, yielding a forgetful functor $\forget:\Pt(\Stoch)^+\to\Stoch^+$. However, the commutativity condition (\ref{eq:ptmap}) must be tweaked to accommodate operations with higher arity:
\begin{equation}\label{eq:ptOperad}\vcenter{\xymatrix{
		1 \ar@{~>}[d]_{r} \ar@{~>}[r]^{r} & X \ar@{~>}[r]^-p  & \coprod_{i} Y_{i} \ar[r]^-{i} &**[r]\coprod_{i} 1\cong n \ar@{~>}[d]^-{\coprod_{i} s_{i}}\\
		X \ar@{~>}[rrr]_p &&&  \coprod_{i} Y_{i}.\\
}}\end{equation}
The lower path $r.p$ weights the kernel $p$ by the prior $r$. The upper path marginalizes $r.p$ to obtain a distribution on the indices $i\in n$, and uses this to weight the prior distributions over $Y_{i}$. Moreover, the top row in (\ref{eq:ptOperad}) defines a functor ${\aggr:\Pt(\Stoch)^+\to\Prob}$ sending a kernel $p:\<Y_{i},s_{i}\>\to\<X,r\>$ to the aggregate distribution $|p|$ shown below:\footnote{The following diagram shows that such kernels are closed under composition:
	\[\xymatrix{
		1 \ar@{~>}[dd]_{r} \ar@{~>}[rd]^{|p|} \ar@{~>}[rrrr]^{r} &&&& X \ar@{~>}[d]^{p} \\
		&  **[r] n\cong\coprod_{i} 1 \ar@{~>}[d]_{\coprod_{i}s_{i}} \ar@{~>}[rrr]^-{\coprod_{i} s_{i}} \ar@{~>}[rrd]^(.6){\coprod_{i}|q_{i}|} &&& \coprod_{i} Y_{i} \ar@{~>}[d]^-{\coprod_{i} q_{i}}\\
		X \ar@{~>}[r]_{p} & \coprod_{i} Y_{i} \ar@{~>}[r]_{\coprod_{i} q_{i}} & \coprod_{ij} Z_{ij} \ar[r]_-{\coprod_{i} j} & \coprod_{i} m_{i}\cong\coprod_{ij} 1 \ar@{~>}[r]_-{\coprod_{ij}\gamma_{ij}} & \coprod_{ij}Z_{ij} \\
	}\]
}
\begin{equation}\label{eq:probproj}\vcenter{\xymatrix{
		1 \ar@{~>}[rrr]^{|p|} \ar@{~>}[rd]_{r} &&& n \\
		& X \ar@{~>}[r]_-{p} & \coprod_{i} Y_{i} \ar[ur]_{i} &.\\
}}\end{equation}

This realizes our goal for the section. A synthesis of the probabilistic models from section \ref{sec:fail} and the possibilistic failure modes of section \ref{sec:failmode} can be modeled as a joint lifting $\SS:\LSI\to\Pt(\Stoch)^+$:
\begin{equation}\vcenter{\xymatrix@!0@R=5ex@C=14ex{
		&& \LSI \ar[lldd]_{\PP} \ar[dd]_{\SS} \ar[rrdd]^{\MM} && \\
		&&&&\\
		\Prob &&\Pt(\Stoch)^+ \ar[r]_-{\forget} \ar[ll]^-{\aggr}&\Stoch^+ \ar[r]_-{\supp} & \Rel^+\\
}}\end{equation}

\section{Conclusion}

We close by noting, briefly, some limitations of the present paper. Though our model is based on a real-world system, it is much too coarse to use in practice. A true predictive model would require a much more detailed decomposition of the system. Similarly, the failure model presented in Section \ref{sec:fail} is not based on real data, but rather selected to illustrate certain points. However, we intend to refine the model over time and, eventually, hope to produce a reference implementation for categorical systems modeling.

Second, the models presented here are purely static, but it would be preferable to incorporate sensor observations into our failure predictions. Here we should be able to leverage existing work on the interpretation of port-graphs as behavioral constraints on dynamical systems. Our next goal is to develop a dynamical model of system and component behaviors including both functioning and malfunctioning components. By substituting malfunctioning component models into the larger dynamical system, we can estimate the likely sensor readings for each failure mode and use these to assess the relativized failure probabilities discussed in Section \ref{sec:failmode}.\\

\noindent\footnotesize \textbf{Disclaimer}:  Commercial products are identified in this article to adequately specify the material. This does not imply recommendation or endorsement by the National Institute of Standards and Technology, nor does it imply the materials identified are necessarily the best available for the purpose.

\normalsize

\nocite{*}
\bibliographystyle{eptcs}
\bibliography{LSI}

\begin{thebibliography}{1}
\providecommand{\bibitemdeclare}[2]{}
\providecommand{\surnamestart}{}
\providecommand{\surnameend}{}
\providecommand{\urlprefix}{Available at }
\providecommand{\url}[1]{\texttt{#1}}
\providecommand{\href}[2]{\texttt{#2}}
\providecommand{\urlalt}[2]{\href{#1}{#2}}
\providecommand{\doi}[1]{doi:\urlalt{http://dx.doi.org/#1}{#1}}
\providecommand{\bibinfo}[2]{#2}

\bibitemdeclare{article}{LSI}
\bibitem{LSI}
\bibinfo{author}{John~S \surnamestart Beers\surnameend} \&
  \bibinfo{author}{William~B \surnamestart Penzes\surnameend}
  (\bibinfo{year}{1999}): \emph{\bibinfo{title}{The NIST length scale
  interferometer}}.
\newblock {\sl \bibinfo{journal}{Journal of Research of the National Institute
  of Standards and Technology}} \bibinfo{volume}{104}(\bibinfo{number}{3}), p.
  \bibinfo{pages}{225}, \doi{10.6028/jres.104.017}.

\bibitemdeclare{book}{LeinsterOperad}
\bibitem{LeinsterOperad}
\bibinfo{author}{Tom \surnamestart Leinster\surnameend} (\bibinfo{year}{2004}):
  \emph{\bibinfo{title}{Higher Operads, Higher Categories}}.
\newblock \bibinfo{series}{London Mathematical Society Lecture Note Series},
  \bibinfo{publisher}{Cambridge University Press},
  \doi{10.1017/CBO9780511525896}.

\bibitemdeclare{article}{operad}
\bibitem{operad}
\bibinfo{author}{David~I \surnamestart Spivak\surnameend}
  (\bibinfo{year}{2013}): \emph{\bibinfo{title}{The operad of wiring diagrams:
  Formalizing a graphical language for databases, recursion, and plug-and-play
  circuits}}.
\newblock {\sl \bibinfo{journal}{arXiv preprint
  \href{https://arxiv.org/abs/1305.0297}{arXiv:1305.0297}}}.

\end{thebibliography}

\newpage

\section*{Appendix}

The appendix collects an explicit description of the $\LSI$ operad discussed throughout the paper. Interface types and boundaries are shown in Table \ref{tab:LSI1}. Operations, along with combinatorial descriptions of the associated port-graph architectures, and an equation are shown in Table \ref{tab:LSI2}.

\begin{table}[h]
	\centering
	\begin{tabular}{c}
		\begin{tabular}{|c|l|}
			\multicolumn{2}{c}{$\mathbf{LSI\ System\ Boundaries}$}\\\hline
			$\mathbf{Boundary}$& $\mathbf{Ports}$\\\hline
			$\Interfer$ & $\laser,\fringe$\\
			$\Chassis$ & $\laser,\focus,\drive$\\
			$\Optics$ & $\focus,\intens$\\
			$\Bath$ & $\heat,\water,\setPt$\\
			$\BoxTemp$ & $\heat_1,\heat_2,\laser,\temp$\\
			$\Lab$ & $\heat,\temp$\\
			$\Mixer$ & $\mix$\\
			$\Res$ & $\heat_1,\heat_2,\mix,\water$\\
			$\Heater$ & $\setPt,\heat$\\
			$\TempSys$ & $\laser,\water,\temp_1,\temp_2,\setPt$\\
			$\LengthSys$ & $\laser,\intens,\fringe,\drive$\\
			$\Sensors$ & $\intens,\fringe,\temp_1,\temp_2,$\\
			& $\focus,\heat,\laser$\\
			$\Actuators$ & $\water,\drive,\setPt,\focus,\heat,\laser$\\
			$\LSIob$ & $\water,\intens,\fringe,\temp_1,$\\
			& $\temp_2,\setPt,\drive$\\\hline
		\end{tabular}\\\\
		$\IType=\left\{
		\begin{array}{l}
		\heat, \laser, \water, \focus, \mix, \temp,\\
		\fringe, \intens, \drive, \setPt\\
		\end{array}
		\right\}$\\
	\end{tabular}
	\caption{Objects and types for the LSI system operad, compiled from Figures \ref{fig:TempSys}, \ref{fig:subst} and \ref{fig:decomps}.}
	\label{tab:LSI1}
\end{table}

\begin{table}[h]
	\centering
	\begin{tabular}{|rl|}
		\multicolumn{2}{c}{\textbf{LSI System Architectures}}\\\hline
		$\varphi$:&$(ls:\LengthSys,ts:\TempSys)\longrightarrow\LSIob$\\
		&$\ell.\laser=t.\laser$\\
		\hline
		$\lambda$:& $(in:\Interfer,op:\Optics, ch:\Chassis)\longrightarrow\LengthSys$\\
		&$ch.\focus=op.\focus$\\
		&$ch.\laser=in.\laser$\\
		\hline
		$\tau$:&$(ba:\Bath,bt:\BoxTemp,rt:\Lab)\longrightarrow\TempSys$\\
		&$bt.\heat_1=ba.\heat$\\
		&$bt.\heat_2=rt.\heat$\\
		\hline
		$\kappa$:&$(sn:\Sensors,ac:\Actuators)\longrightarrow\LSIob$\\
		&$s.\heat=a.\heat$\\
		&$s.\laser=a.\laser$\\
		&$s.\focus=a.\focus$\\
		\hline
		$\sigma$:&$(rt:\Lab,bt:\BoxTemp,op:\Optics,ls:\Interfer)\longrightarrow\LengthSys$\\
		&$bt.\heat_2=rt.\heat$\\
		&$bt.\laser=ls.\laser$\\
		\hline
		$\alpha$:&$(ch:\Chassis,ba:\Bath)\longrightarrow\Actuators$\\
		&No equations\\
		\hline
		$\beta$:&$(ht:\Heater,mx:\Mixer,rs:\Res)\longrightarrow\Bath$\\
		&$rs.\mix=mx.\mix$\\
		&$rs.\heat_2=ht.\heat$\\
		\hline
		\multicolumn{2}{c}{}\\
		\multicolumn{2}{c}{\textbf{Architectural Coherence Equation}}\\
		\multicolumn{2}{c}{$\varphi(\lambda,\tau)=\kappa(\sigma,\alpha)$}\\
	\end{tabular}
	\caption{Architectures (operations) from the LSI system operad. Each architecture can be described as a set of equations between component ports. The architectural coherence equation corresponds to the diagram in Figure~\ref{fig:decomps}.}
	\label{tab:LSI2}
\end{table}

\end{document}